\documentclass[twocolumn,showpacs, secnumarabic,amssymb, nobibnotes, aps, prd]{revtex4}

\usepackage{graphicx}
\usepackage{dcolumn}
\usepackage{bm}

\begin{document}

\title{Transverse Field Ising Ferromagnetism in Mn$_{12}$-acetate-MeOH}
\author{P. Subedi}
\affiliation{Department of Physics, New York University, New York, New York 10003, USA}
\author{Bo Wen}
\affiliation{Department of Physics, City College of New York, CUNY, New York, New York
10031, USA}
\author{Y. Yeshurun}
\affiliation{Department of Physics, Institute of Nanotechnology, Bar-Ilan University,
Ramat-Gan 52900, Israel}
\author{M. P. Sarachik}
\affiliation{Department of Physics, City College of New York, CUNY, New York, New York
10031, USA} 
\author{\\A. D. Kent}
\affiliation{Department of Physics, New York University, New York, New York 10003, USA}
\author{A. J. Millis}
\affiliation{Department of Physics, Columbia University, New York, New York 10027, USA}
\author{S. Mukherjee}
\author{G. Christou}
\affiliation{Department of Chemistry, University of Florida, Gainesville, Florida 32611,
USA}

\begin{abstract}
We report measurements of the magnetic susceptibility of single crystals of Mn$_{12}$-acetate-MeOH, a new high-symmetry variant of the original single molecule magnet Mn$_{12}$-acetate. A comparison of these data to theory and to data for the  Mn$_{12}$  acetate material  shows that Mn$_{12}$-acetate-MeOH is a realization of a transverse-field Ising ferromagnet in contrast to the original Mn$_{12}$ acetate material, in which solvent disorder leads to effects attributed to random field Ising ferromagnetism.  
\end{abstract}

\maketitle

Dipolar interactions that can lead to long range magnetic order have been explored extensively for decades, both theoretically \cite{PhysRev.70.954} and experimentally \cite{PhysRevLett.65.1064}. Recent interest in dipolar magnetism has focused on 
systems in which quantum fluctuations of the spins compete with the dipolar long range order.  Of particular interest in this context are the single molecule magnets (SMM) composed of molecules of transition-metal ions that behave as single, rigid magnetic entities with a large ground-state spin and a strong uniaxial anisotropy, leading to Ising-like behavior. The distance between molecules in this SMM is sufficiently large that the dominant interaction is the dipolar interaction. For spins on an ordered lattice, dipolar interactions lead to a ground state with long ranged order which may (depending on the lattice structure) be ferromagnetic or antiferromagnetic \cite{PhysRevLett.90.017206,PhysRevLett.93.117202,LuisPRL2005,evangelisti:167202,FernandezPRB2000,Millis09,garanin:174425}. However, the application of a magnetic field in a direction transverse to the Ising axis induces quantum spin fluctuations that compete with the long-range order by mixing the eigenstates of $S_z$ \cite{FriedmanJAP1997}. This interplay between the long range order and spin fluctuations is described by the Transverse-Field Ising Hamiltonian:
\begin{equation}
 \mathcal{H}= \frac{1}{2} \sum_{i\neq j} J_{ij} S^z_i S^z_j+  \Delta \sum_iS_i^x
 \label{TFI}
 \end{equation}
Here, $S_i$ is a two level Ising spin on lattice site $i$, $J_{ij}$ are the dipolar couplings and $\Delta$ is the tunnel splitting that depends on the applied transverse field \cite{Millis09}. This Hamiltonian applies at energies and temperatures such that excitation to higher energy states of the molecular  complex can be neglected; for the systems of interest in this paper Eq.~\ref{TFI} applies below   a transverse-field-dependent temperature $\lesssim 6$ K.
 
One of the most studied SMM is Mn$_{12}$-ac. In this compound the Mn$_{12}$ units  crystallize in a body centered tetragonal lattice and are well separated by acetate solvent molecules. Each Mn$_{12}$ molecule behaves as a nanomagnet with spin $S = 10$ oriented along the crystallographic \textit{c} axis by a strong anisotropy, $DS_z^2 \approx 65$ K \cite{JonathanPRL1996}. For this lattice structure the ground state of Eq.~\ref{TFI} at $\Delta=0$ is ferromagnetically ordered, so we refer to the system as a Transverse Field Ising Ferromagnet (TFIFM). However, Mn$_{12}$-ac is not simply a representation of the transverse-field Ising model because a distribution in the arrangements of the solvent molecules results in a distribution of discrete tilts of the molecular magnetic easy axis from the global (average) easy axis of a crystal, thereby locally breaking the global tetragonal symmetry of the crystal  \cite{PhysRevLett.91.047203,PhysRevLett.90.217204,PhysRevB.70.094429,AndyJLTP2005}.  Although the small molecular easy-axis tilts  ($\approx \pm 1^\circ$) induce only minor perturbations in the dipolar interaction, an external transverse magnetic field has projections along (the randomly distributed) easy axes that become comparable in magnitude to the dipolar field itself for transverse field magnitudes of order $4$ T \cite{Millis09, BoPRB2010}. It was recently shown \cite{Millis09, BoPRB2010} that one can account for the experimental data for Mn$_{12}$-ac by adding a site transverse field-dependent random-field term $\sum_{i} h_{i} S^z_i$ to Eq.~\ref{TFI} so that this prototypical molecular magnet is a realization of the Random-Field Ising Ferromagnet (RFIFM).

In this paper we report results of an investigation of [Mn$_{12}$O$_{12}$(O$_2$CMe)$_{16}$(MeOH)$_4$]$\cdot$MeOH, hereafter referred to as  Mn$_{12}$-ac-MeOH, a new high-symmetry variant of the original single molecule magnet Mn$_{12}$-ac.  The two systems differ only in the isomer disorder introduced by the solvent molecules in Mn$_{12}$-ac, so that a comparison of their magnetic response provides quantitative information about the effect of random fields. We find that the behavior of Mn$_{12}$-ac-MeOH is consistent with Eq.~\ref{TFI} without disorder effects at intermediate temperatures, where the new ``pure'' MeOH variant represents a model-system for the study of intrinsic transverse-field Ising magnetism. However, deviations from simple theoretical expectations for both Mn-12 variants below about 2 K are not currently understood and require further study.

Mn$_{12}$-ac-MeOH crystallizes in the space group $I\bar{4}$ with unit cell parameters $a = b = 17.3500(18)$ \AA, $c = 11.9971(17)$ \AA, molecules per unit cell (Z) = 2, $V = 3611.4$ \AA$^3$ at $-100\,^{\circ}\mathrm{C}$ \cite{Stamatatos,PhysRevB.80.094408, Bian2005}, unit-cell spin (S = 10) and anisotropy (D = -0.667 K) is nearly identical to Mn$_{12}$-ac (space group $I\bar{4}$; unit cell parameters $a = b = 17.1668(3)$ \AA, $c = 12.2545(3)$ \AA, $Z = 2, V = 3611.39$ \AA$^3$ at $83$ K) \cite{corniaACC2002}. However, as described in detail in Ref.~\cite{PhysRevB.80.094408}, unlike its close relative Mn$_{12}$-ac, the local symmetry associated with the solvent of crystallization retains the overall molecular $S_4$ global symmetry of the crystal.

\begin{figure}[tbh]
\centering
\includegraphics[width=\linewidth]{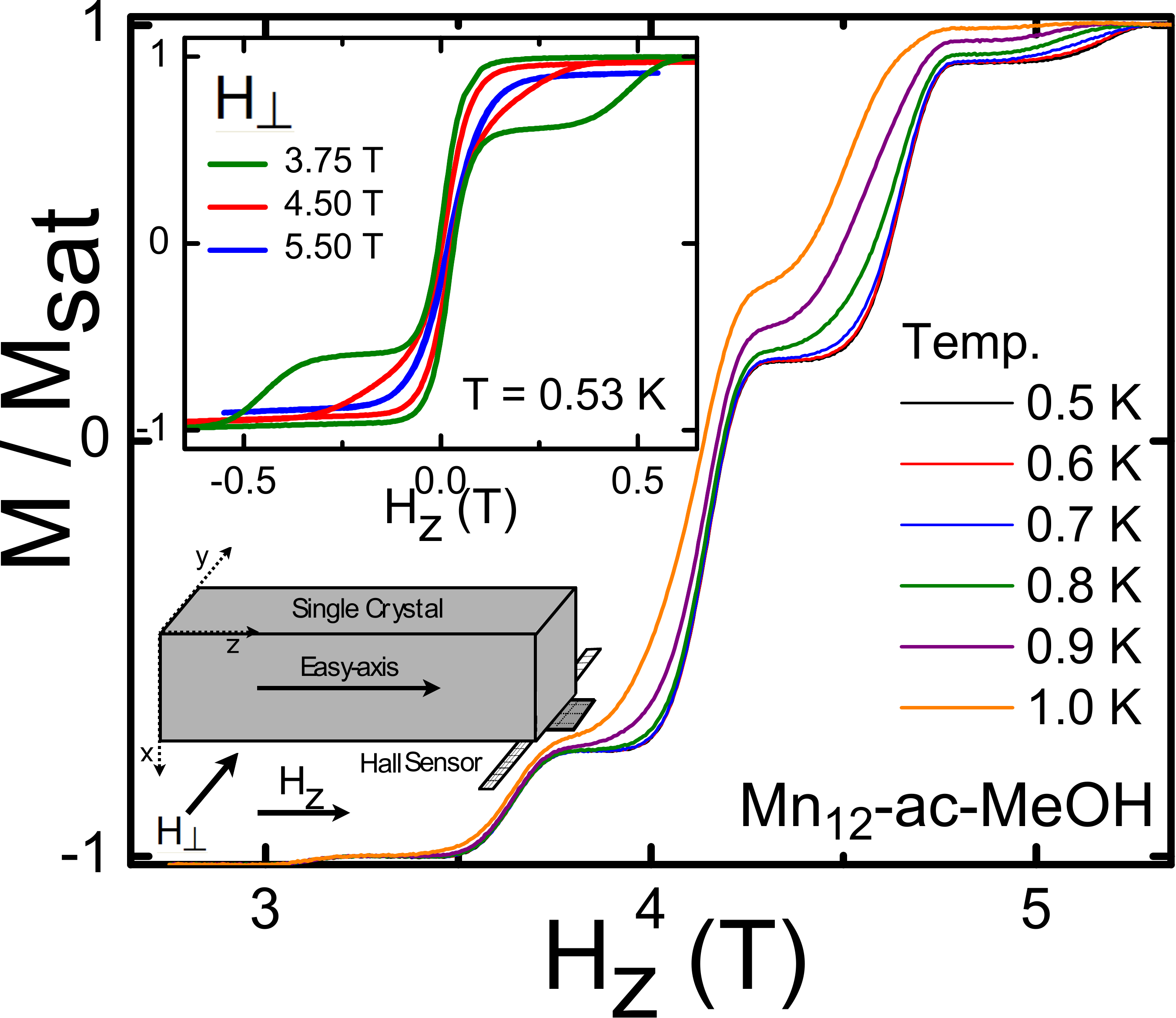}
\caption{(Color on line) Normalized magnetization of Mn$_{12}$-ac-MeOH as a function of longitudinal magnetic field, $H_z$, in zero transverse field at several temperatures below $1$ K.  Top Inset: Magnetization vs  $H_z$ at $T = 0.53$ K for different transverse fields, $H_\perp$.  Bottom Inset: Schematic diagram of the sample, the Hall sensor and magnetic fields.}
\label{steps}
\end{figure}

Measurements of the longitudinal magnetization and susceptibility were performed on three Mn$_{12}$-ac-MeOH single crystals of dimensions $\sim 0.2 \times 0.2 \times 0.95$ mm$^3$, $0.085 \times 0.085 \times 0.68$ mm$^3$, and $0.075 \times 0.075 \times 0.85$ mm$^3$ (samples A, B and C, respectively). The samples were coated with Paratone $^{\textregistered}$ N to prevent degradation by crystal lattice desolvation \cite{PhysRevB.80.094408}. A Hall sensor, (active area $20 \times 100\ \mu$m$^2$) was used to measure the magnetization, $M_z$, along the easy direction (\textit{c}-axis) of the crystal via measurement of the stray field $B_x$, which is a linear function of $M_z$.  Care was taken to align the sample and the Hall array (placed in the y-z plane) relative to each other and relative to the magnet axes. The relative position between the crystal and the Hall sensor array is shown in the bottom inset of Fig. \ref{steps}; the sensor is placed nearest to the end of the sample, since the stray field, $B_x$, is largest near the edge. Preliminary data for sample A, not corrected for demagnetization effects, is presented in Ref \cite{BoArXiv2010}. For comparison, we also show results of similar measurements for three Mn$_{12}$-ac crystals of dimensions $\sim 0.4 \times 0.4 \times 2.17$ mm$^3$, $\sim 0.4 \times 0.4 \times 2.4$ mm$^3$ and $0.3 \times 0.3 \times 1.85$ mm$^3$ (crystals D, E and F, respectively). All measurements were taken between $0.5$ K and $6$ K in a $^3$He refrigerator in a 3D vector superconducting magnet. A longitudinal field, $H_z$, was swept along the sample's easy axis at rates between $1 \times 10^{-5}$ T/s and $6.7 \times 10^{-3}$ T/s, in the presence of a series of fixed transverse fields, $H_\perp$ (up to 6.8 T) applied in the $y$ direction (see bottom inset of Fig. \ref{steps}). The point labeled $H_z=0$ was determined by symmetry from full hysteresis loops taken between $-0.7$ and $0.7$ T.

The field dependence of the magnetization of Mn$_{12}$-ac-MeOH (Sample C) is shown in the main panel of Fig. \ref{steps} for temperatures below $1$ K in the absence of  transverse magnetic field. Characteristic of resonant tunneling in molecular magnets, the steps occur due to faster spin-reversal at specific (temperature-independent) magnetic fields corresponding to energy-level coincidences on opposite sides of the anisotropy barrier \cite{JonathanPRL1996}. The resonant fields at which the steps occur in Mn$_{12}$-ac-MeOH are the same as in Mn$_{12}$-ac, indicating that the two systems have similar spin energy-level structures. The magnetization exhibits hysteresis due to slow relaxation below a blocking temperature, $T_B$, that depends on the rate at which the magnetic feld is swept. Equilibrium can be established by increasing the temperature and/or decreasing the sweep rate. It can also be promoted by applying a transverse magnetic field. The latter is demonstrated in the top inset of Fig. 1 which shows the low-field magnetization in transverse field at 0.53 K. While hysteresis is evident for H$_{\bot}$ = 3.75 T, the system is in equilibrium at the higher field of H$_\bot$ = 5.5 T; by mixing the eigenstates of S$_z$, the transverse field promotes quantum tunneling and accelerates relaxation toward equilibrium. Care was taken to ensure that the susceptibility and magnetization reported and discussed in the remainder of this paper are equilibrium values. 

Under equilibrium conditions, the longitudinal magnetic susceptibility, $\chi\equiv\partial M_z/\partial H_z|_{H_z=0}$, can be deduced from the slope of $M_z$ versus $H_z$ at $H_z=0$ as described in Ref. \cite{BoPRB2010}.  
\begin{figure}[tbh]
\centering
\includegraphics[width=\linewidth]{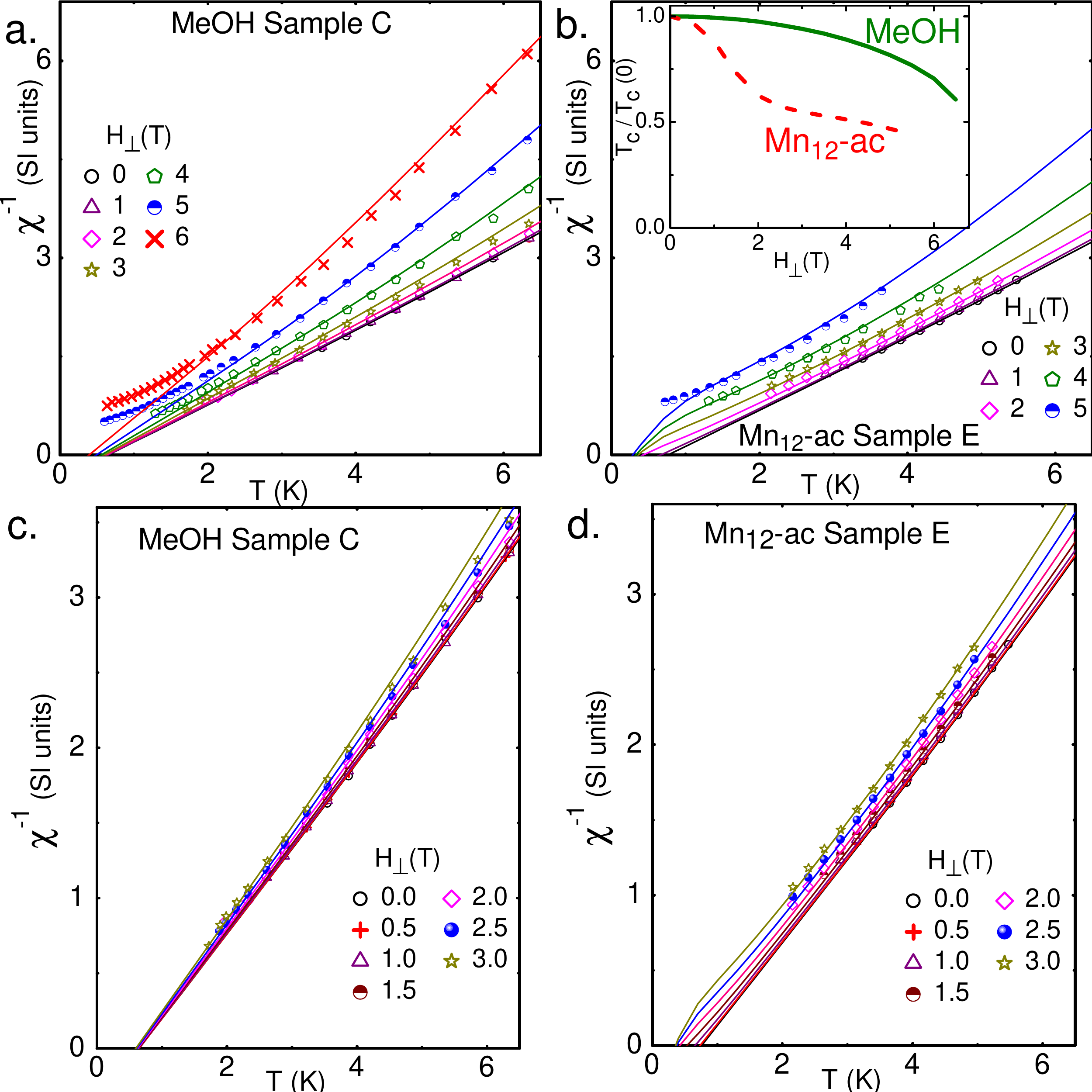}\caption{(Color on line)(a): Inverse susceptibility as a function of temperature for Mn$_{12}$-ac-MeOH (Sample C) in various transverse fields up to $6$ T. (b): Inverse susceptibility as a function of temperature of Mn$_{12}$-ac (Sample E) for various transverse fields up to $5$ T.  (c) and (d) Inverse susceptibility in low transverse field up to 3 T for  Mn$_{12}$-ac-MeOH and Mn$_{12}$-ac, respectively.  The solid lines are theoretical curves based on Eq. 2.  Inset to frame (b): The Weiss temperature $T_{W}(H_\perp)$, normalized to $T_W$ in zero transverse field, for Mn$_{12}$-ac-MeOH (solid green) and for Mn$_{12}$-ac (dashed red) obtained from fits to the theory in the range $2-6$ K.  The dashed lines are the theoretical curves based on Eq. 2 with $\theta=1.8\,^{\circ}$.}
\label{fig2}
\end{figure}
Figures 2(a) and 2(c) show the inverse susceptibility of a Mn$_{12}$-ac-MeOH (Sample C) crystal as a function of temperature for various fixed transverse magnetic fields between 0 and 6 T. The data were corrected for demagnetization effects, as outlined in the Appendix.  The inverse susceptibility increases with transverse field (the susceptibility $\chi$ decreases) due spin canting.  An unexpected flattening of the curve occurs for temperatures below $\sim 2$ K.  For comparison, we show similar data obtained for a Mn$_{12}$-ac (Sample E) crystal in Figs. 2(b) and 2 (d); the expected overall decrease of the susceptibility with transverse field is also observed, as well as the anomalous deviations at low temperature.  On the other hand, it is clear that the response to transverse field is distinctly different for the two systems: while the slopes of the $\chi^{-1}$ vs. $T$ curves increase rapidly for Mn$_{12}$-ac-MeOH, the curves remain approximately parallel with little change of slope in the case of Mn$_{12}$-ac, with a concomitant rapid decrease of the apparent intercept and Weiss temperature, as reported earlier for this random system \cite{BoPRB2010}.

To further demonstrate  the different response to the magnetic field, and guided by  Eq.~\ref{Millis32} below, we plot in Fig. 3 the normalized change  $\Delta \chi^{-1}(H_\perp)$  of the inverse susceptibility at a particular temperature, as a function of $H_\perp^2$ for three Mn$_{12}$-ac-MeOH samples (green dots) and three Mn$_{12}$-ac (red squares) at T = 3.2 K.  We note that the subtraction, $\Delta \chi^{-1}(H_\perp) = \chi^{-1}(H_\perp) - \chi^{-1} (0)$, eliminates the intermolecular interaction term (J), and the normalization removes the dependence on sample volume. Figure 3 clearly shows that the effect of the transverse field is much larger for Mn$_{12}$-ac.

To analyze the data quantitatively we turn to the theoretical expression presented in Ref.~\onlinecite{Millis09}:
\begin{equation}
\chi^{-1}(H_\bot,T)=C\left(\sec^2\theta(H_\bot)\frac{\Delta(H_\bot)}{\tanh\frac{\Delta(H_\bot)}{T}} -J\right).
\label{Millis32}
\end{equation}
Here $J$ is the effective exchange interaction obtained from the appropriate spatial average over the dipole interaction, the angle $\theta$ characterizes the spin canting in an applied transverse field and $\Delta$ is the tunnel splitting; detailed expressions for the  dependence of $\theta$ and $\Delta$ on $H_\bot$ are given in Ref.~\onlinecite{Millis09}.  The bottom line is that the tunnel splitting becomes non-negligible only for $H_\bot>6T$ while in the range $0<H_\bot<5T$,  $\theta~[rad]\approx 0.1H_\bot~[T]$. 

\begin{figure}[tbh]
\centering
\includegraphics[width=\linewidth]{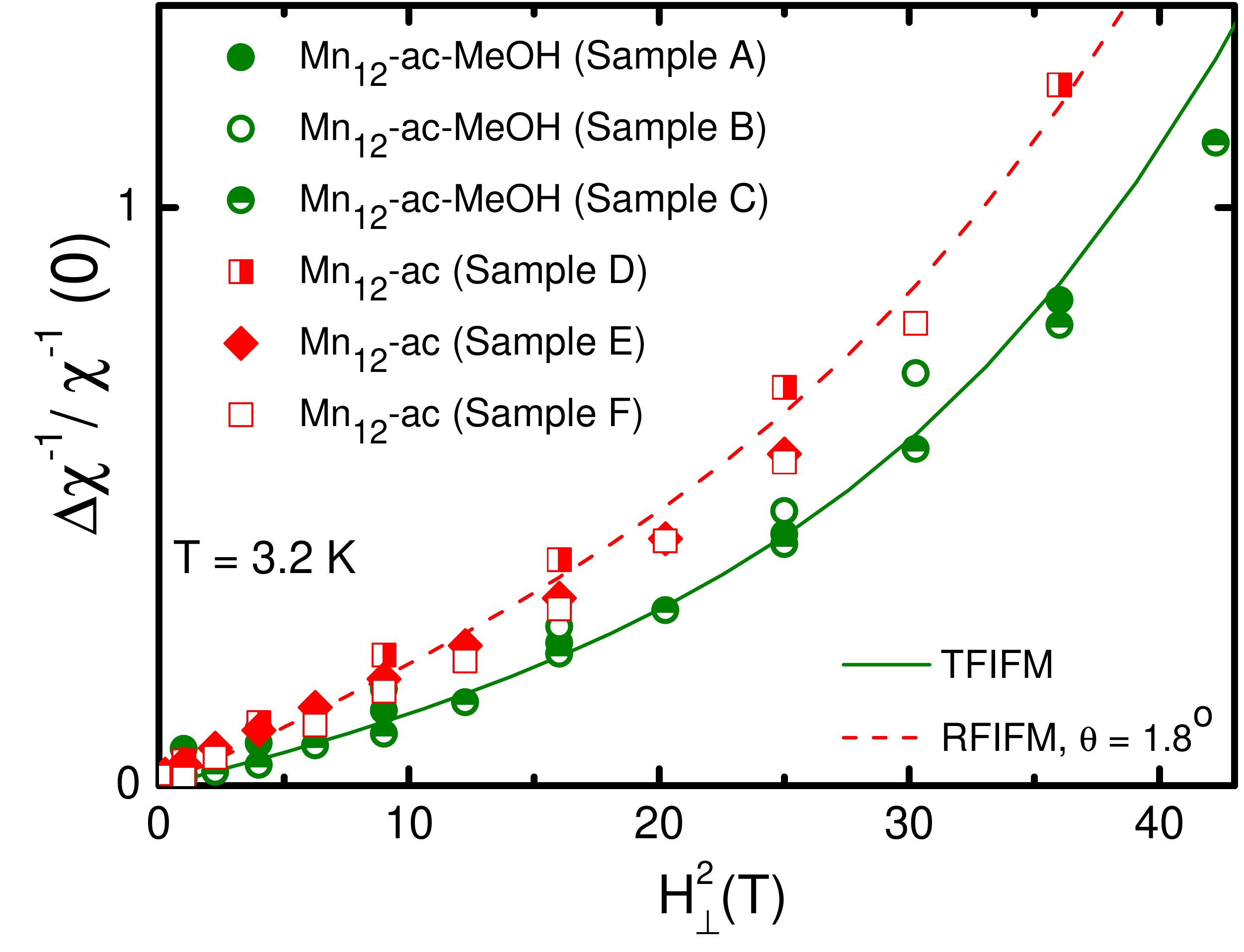}
\caption {(Color on line) The change in inverse susceptibility, $\Delta\chi^{-1}$, normalized to the susceptibility at zero field versus $H_\perp^2$ for Mn$_{12}$-ac-MeOH (green dots, open circle Ref[21]) and Mn$_{12}$-ac (red squares, half filled and solid squares are Sample A and B in Ref[16]) at $T = 3.2$ K. The dashed and dashed dotted lines are calculated using the random-field model of \cite{BoPRB2010} (RFIFM) for two different root mean square tilt angles, as indicated. The solid green line shows the result for the case with no tilt angle (TFIFM).}
\label{fig3}
\end{figure}

Using Eq. 2, we plot in Fig. \ref{fig3}  the change  $\Delta \chi^{-1}$  of the inverse susceptibility, normalized to the zero transverse field inverse susceptibility value,
\begin{equation}
\Delta\chi^{-1} (H_\perp)=\frac{\chi^{-1}(H_\perp)-\chi^{-1}(0)}{\chi^{-1} (0)},
\end{equation}
for the ``pure'' Transverse Field Ising Ferromagnet with no tilt angle (solid line) and for the Random Field Ising Ferromagnet (RFIFM) with two different root mean square tilt angles of $1.8\,^{\circ}$ and $3\,^{\circ}$ (dotted-dashed and dashed lines, respectively) for the random-field distribution proposed by Park et. al ~\cite{Park04}.  The excellent agreement between calculation and data for the MeOH material at $H_\bot<4T$ at $3.2$ K  is an indication that this system is a realization of the dipolar Ising model in a transverse field.  A good fit is obtained for the Mn$_{12}$-ac crystal data with the RFIFM model using root mean square tilt angles of $1.8\,^{\circ}$.

The different amount of disorder in the two systems demonstrated in Fig. \ref{fig3} is also reflected in the temperature dependence of the susceptibility. The theoretical predictions for $\chi^{-1}(T)$ for a pure Mn$_{12}$ system and for a system with an average tilt angle of $1.8\,^{\circ}$ are shown by the solid lines in Fig. 2(a) and 2(b), respectively. The demagnetization correction is obtained by requiring that theory and experiment coincide at zero field. Within this assumption, the data between 2 and 6 K are consistent with theory for both samples, where a particularly good fit is obtained for fields below 3 T. However, while the theoretical lines intersect the temperature axis at T$_{W}$(H$_\bot$) implying the approach to a ferromagnetic phase, the measured susceptibility deviates from this simple behavior, flattening as the temperature decreases toward the presumed transition.   The behavior observed at these low temperatures is not understood, and may imply that a transition to a new phase is prohibited for reasons that are unclear.  It is nevertheless interesting to examine the Weiss temperatures T$_{W}$(H$_\bot$) predicted by the theory.  This is shown in the inset to Fig. 2(a)  for both samples, based on fits of the susceptibility measured between 2 and 6 K, where the ``phase diagram'' for the pure case is denoted by the green solid line and the red dashed line denotes the theoretical prediction for the disordered case with average tilt angle of $\theta=1.8\,^{\circ}$.  For the theoretical calculation, the dipolar part of the interaction was obtained using the measured lattice parameters and crystal structure of  Mn$_{12}$-ac-MeOH  and the spin canting and tunnel splitting were obtained as described in Ref.~\onlinecite{Millis09}.  A phase diagram similar to that observed for Mn$_{12}$-ac-MeOH was obtain by Burzuri {\it et al.} in Fe$_8$ \cite{Burzuri}. 

The initial  suppression of $T_{W}$ for transverse fields $H_\bot<5$ T (see inset to Fig. 2(a)) is expected due to spin canting, which reduces the net moment in the axial direction; the more rapid suppression at higher fields derives from the tunnellng term. A substantially more rapid suppression of $T_{W}$ with $H_\bot$ is evident for Mn$_{12}$-ac. The results for Mn$_{12}$-ac are consistent with a modified theory that includes the effects of random fields arising from the tilt angles.

In summary, these studies demonstrate that the magnetic susceptibility of Mn$_{12}$-ac-MeOH follows the behavior expected for a transverse field Ising ferromagnet, in marked contrast with Mn$_{12}$-ac.  The temperature dependence of the susceptibility and the dependence of the (extrapolated) Weiss temperature on applied transverse field are different for the two materials. More broadly, the availability of these two very similar SMMs with distinct types of magnetism provides unique opportunities for experimental studies of the effect of randomness on quantum phase transitions and magnetic relaxation. In particular, large transverse fields ($\sim 5$ T) that enhance pure quantum tunnel relaxation (tunneling relaxation without the need for thermal activation) will enable equilibrium susceptibility studies down to very low temperature (mK).   Such investigations may reveal interesting ground states (ferromagnetic, spin glass, or even antiferromagnetic) that may differ for the pure and random systems.

We thank A. Narayanan for technical assistance during the experiment. Support for GC was provided under grant CHE-0910472; ADK acknowledges support by ARO W911NF-08-1-0364 and NYU-Poly Seed Funds; MPS acknowledges support from NSF-DMR-0451605; YY acknowledges support of the Deutsche Forschungsgemeinschaft through a DIP project. AJM acknowledges support of NSF-DMR-1006282.

\section{Appendix}

\vspace{0.05in}
In this Appendix we discuss the procedure used to convert the experimental measurements to the susceptibilities reported in the paper. 

Our data at zero and nonzero transverse field are obtained with a Hall sensor that reads out a Hall voltage proportional to the component of the magnetic field perpendicular to the sample surface at the position of the sensor.  This magnetic field component is perpendicular to the applied magnetic field and is generated by the sample magnetization in a way that depends on the sample shape and spatial distribution of the magnetization.   The proportionality constant is not known $a priori$.  Also, the Hall bar must be placed at the edge of the sample, where the demagnetization correction may be significant but is not easy to model.  Determining the `intrinsic' sample-averaged magnetization from the Hall data requires analysis.
 
Theory \cite{Millis09} indicates that at temperatures sufficiently high relative to disorder scales and to  the ordering temperature, but low enough relative to the energy of the next spin manifold, the susceptibility $\chi$ may be written
\begin{equation}
\chi^{-1}=A\left(\frac{T}{cos^2 \theta_H}-J^{*}\right)
\label{chiinvtheory}
\end{equation}
where the amplitude $A$ is related to sample volume and spin magnitude, $cos\theta_H$ expresses the tilting of spins in a field transverse to the Ising axis, and $J^{*}$ is an effective exchange constant that includes a significant correction due to the demagnetization field in the case of dipolar magnets such as Mn$_{12}$. In the absence of an applied transverse field, Eq.~\ref{chiinvtheory} with a $J^{*}=0.74K$ describes the temperature dependence measured by a SQUID magnetometer over a wide temperature range, extending from the blocking temperature up to a scale $\sim 6K$ above which thermal excitation to other spin manifolds begins to become important.  The intrinsic susceptibility is obtained from SQUID-based measurements of a series of samples with different aspect ratios, extrapolated to the limit of an infinitely long sample for which, for which the demagnetization fields are negligible \cite{ShiqiPRB2010}.

\begin{figure}[tbh]
\centering
\includegraphics[width=\linewidth]{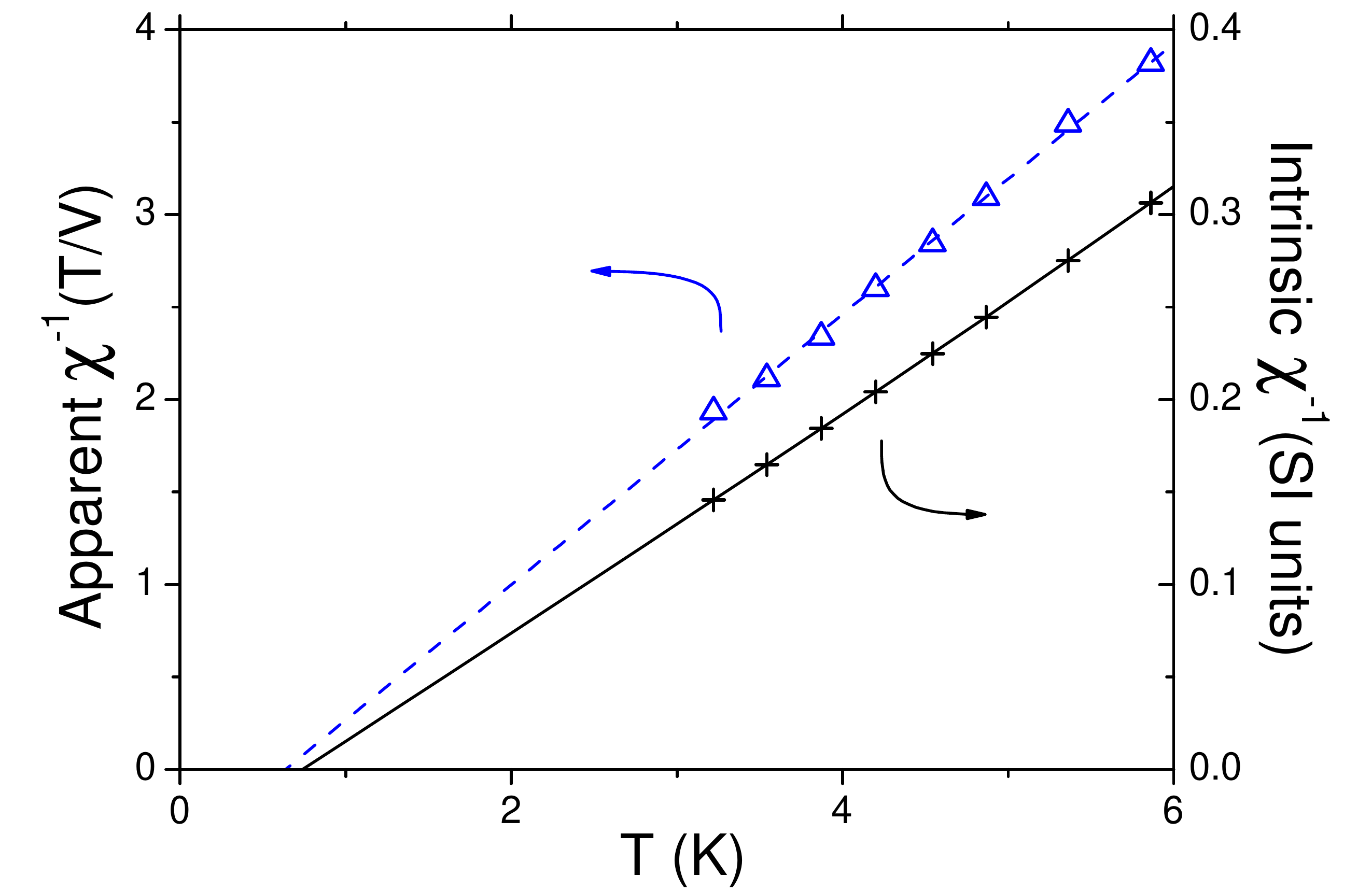}
\caption {(Color on line) Temperature dependence of the intrinsic (crosses) and apparent (triangles) inverse susceptibility of Mn$_{12}$-ac-MeOH for zero transverse field obtained from global SQUID-based measurements \cite{ShiqiPRB2010} and local Hall-sensor measurements (sample C), respectively. Solid and dashed lines are linear fits to the experimental data.}
\label{app}
\end{figure}

From the theoretical results we expect that the susceptibility measured from the Hall bar, as a function of temperature $T$, transverse field $H_T$ and 
\begin{equation}
\chi^{-1}(T,H_T,z)=A\left(T\frac{f(T,H_T)}{cos^2\theta_H}-J^{*}-J_{demag}(z)\right)
\label{chidata}
\end{equation}
Here $A$ is an amplitude relating to the details of the measurement, $J_{demag}$ expresses the local demagnetization effects at the position $z$ of the Hall bar sensor as well as global effects relating to the sample shape, and  $f(T,H_T)$ parametrizes the effects of interactions beyond mean field theory, disorder and transverse fields (apart from the spin tilt effects included in the $cos\theta_H$). Explicit expressions for $f$ obtained within mean field theory are given in Ref.~\cite{Millis09}. It is important that neither $A$ nor $J_{demag}$ is expected to depend on transverse field. 

To analyze the Hall bar  data we compare the Hall bar measurements with ``global''  measurements obtained in a commercial Quantum Design SQUID-based MPMS magnetometer \cite{ShiqiPRB2010} taken at zero transverse field. Figure \ref{app} shows the inverse of the  intrinsic susceptiblity in SI units (crosses and solid line, black on line) appropriate for a crystal with no demagnetization effects, obtained as described in Ref. \cite{ShiqiPRB2010}  by extrapolating SQUID measurements on a series of  MeOH crystals to the limit of infinite aspect ratio.  Representative data derived from the voltage measured by the Hall probe, in units of T/V, are shown for sample C (open triangles and dashed line, blue on line). These data are seen to be linear in temperature and to  intercept the temperature axis at $T = 0.63$ K, a value smaller than $T_c$ due to local demagnetization effects.   We determine the prefactor $A$ from the slope of independently measured SQUID magnetometer data.  We then fix $J_{demag}(z)$ by shifting $\chi^{-1}$ along the temperature axis so that the measured $\chi^{-1}$ extrapolated to zero temperature coincides with the SQUID data. Finally, we use these values of $A$  and $J_{demag}$ determined from our $H_T=0$ analysis to correct the $\chi^{-1}$ at all values of transverse field, $H_T$. In essence, this procedure yields a value for the function $f(T,H_T)$ which expresses the non-mean-field non-pure-system physics, thereby enabling a cross-comparison between different samples and between experiment and theory \cite{Millis09}.

For more information the reader is referred to \cite{BoDemagPaper2011}.


\end{document}